\documentstyle [aaspp4]{article}
\begin{document}
\title{The Structure of the Outer Halo of the Galaxy and its Relationship to Nearby Large-Scale Structure}
\author{F.D.A. Hartwick}
\affil{Department of Physics and Astronomy, \linebreak University of Victoria, Victoria, 
BC, Canada, V8W 3P6}
\begin {abstract}
We present evidence to support an earlier indication that the Galaxy is embedded in an extended, highly inclined, triaxial halo outlined by the spatial distribution of companion galaxies to the Milky Way. Signatures of this spatial distribution are seen in 1) the angular variation of the radial-velocity dispersion of the companion galaxies, 2) the spatial distribution of the M~31 sub-group of galaxies, 3) the spatial distribution of the isolated, mainly dwarf irregular, galaxies of the Local Group, 4) the velocity anisotropy quadrupole of a sub-group of high-velocity clouds, and 5) the spatial distribution of galaxies in the Coma-Sculptor cloud. Tidal effects of M~31 and surrounding galaxies on the Galaxy are not strong enough to have affected the observed structure. We conclude that this distribution is a reflection of initial conditions. A simple galaxy formation scenario is proposed which ties together the results found here with those of Holmberg (1969) and Zaritsky et al. (1997) on the peculiar distribution of satellites around a large sample of spiral galaxies.
\end{abstract}
\section {Introduction}
Our present view of the Milky Way had its origin over 80 years ago with the classic work of Shapley (1918). The `Shapley Galaxy' had the Sun offset from the center in a disk embedded in a 
roughly spherical halo of globular clusters. At the time of Shapley's work the only available probes of the outer halo were globular clusters. It was not until  20 years later that the first dwarf spheroidal companion to the Galaxy was discovered, also by Shapley. A large increase in the discovery of distant low surface brightness satellites came from plates taken with the 48-inch Palomar Schmidt telescope in the early 1950's and has continued up to the present time. It was then not until the mid 1970's that almost simultaneously Lynden-Bell (1976) and Kunkel \& Demers (1976) noted that many of the outlying satellites lay in a plane nearly coincident with the Magellanic Stream of neutral hydrogen. Since then other streams have been discussed (c.f. Lynden-Bell 1982, Majewski 1994, Lynden-Bell \& Lynden-Bell 1995). In what follows we consider this problem from a different perspective, clarify and extend our previous work, (Hartwick 1996a,b, hereafter H96a,b), and reach similar conclusions using an updated sample and an improved  method of data analysis. There it was argued that our `Shapley Galaxy' is embedded in an elongated, highly inclined outer halo defined by the spatial distribution of the associated companion galaxies to the Milky Way. We strengthen this picture by showing that a very similar spatial distribution applies to the M~31 sub-group of galaxies and to the Local Group as defined by the outer isolated, predominately dwarf irregular, galaxies. In addition, we conclude, on the basis of an apparent similarity in these spatial distributions with nearby larger-scale structure, that the results may be understood as a reflection of initial conditions.

\section{Defining the Outer Halo of the Galaxy}

\subsection{Sample Selection}

Any method of defining the limits of the outer halo will be somewhat arbitrary and for this reason we look for common properties within different subsamples of both nearest companion galaxies and globular clusters with Galactocentric distances between 25 kpc and 450 kpc. Within these limits are 10 companion galaxies (treating the LMC and SMC as one system) and 15 globular clusters. The data on the following satellite galaxies were taken from the recent review by Mateo (1998): LMC, Fornax, Leo I, Sculptor, Sextans, Leo II, Ursa Minor, Carina, Draco and Phoenix. The globular cluster data were taken from the most recent web-based update of the compilation of Harris
(1996) and included the following clusters: AM-1, Eridanus, Pal 2, NGC2419, Pyxis, Pal 3, Pal 4, AM-4, NGC5694, NGC5824, Pal 14, NGC6229, Pal 15, NGC7006, and Pal 13.  No radial velocities were available for Pyxis or AM-4 and an `average' value of 16 km sec$^{-1}$ was assumed for Phoenix.
\subsection{The Spatial Distribution}
Whereas in the previous work (H96a,b) the spatial distribution was determined by fitting an ellipsoid to the data by least squares, in this work we have characterized the spatial structure by a symmetric tensor with average values $\overline{x_{i}x_{i}/d_{i}}, \overline{y_{i}y_{i}/d_{i}}, \overline{z_{i}z_{i}/d_{i}}, \overline{x_{i}y_{i}/d_{i}}, \overline{x_{i}z_{i}/d_{i}}, \overline{y_{i}z_{i}/d_{i}}$ as components, where $d_{i}$ is the distance of the i$^{th}$ object from the origin (in this case the Galactic center with R$_{0}$ = 8 kpc assumed) and $x_{i}, y_{i},$ and $z_{i}$ are the projections of $d_{i}$ along the principal axes of the Galactic coordinate system. Eigenvalues of the matrix were obtained in the standard way and are given below along with their directions for various subsamples of the data. Throughout this paper the quoted one sigma errors were computed by the bootstrap method. The errors on quantities involving a direction (i.e. the spatial and radial velocity eigenvalues), were computed in two ways. The first allowed for errors due to differences in direction (as below) while the second assumed a fixed direction and projected the resulting dispersions onto the principal axes. The errors computed by the second method were found to be so similar to those from the first that we have not given them here. The results of these procedures are given below where the units of the eigenvalues (e$_{a,b,c}$) are kpc, and $l$ and $b$ are Galactic coordinates expressed in degrees.

\begin{center}
All Galaxies\ (n=10)
\end{center}
\begin{eqnarray}
e_{a} = 107^{+35.0}_{-35.9} & l = 301^{+49.1}_{-38.4} & b = -76.5^{+17.5}_{-1.30} \nonumber \\
e_{b} = 41.2^{+10.9}_{-14.6} & l = 64.9^{+9.48}_{-3.14} & b = -7.60^{+20.6}_{-16.5} \\
e_{c} = 3.74^{+1.54}_{-1.12} & l = 336^{+8.28}_{-3.79} & b = 11.1^{+3.89}_{-5.05} \nonumber 
\end{eqnarray}

\begin{center}
Galaxies $-$ Phoenix\ (n=9)
\end{center}
\begin{eqnarray}
e_{a} = 74.2^{+21.0}_{-19.7} & l = 15.3^{+29.2}_{-67.3} & b = -74.0^{+20.8}_{-2.48} \nonumber \\
e_{b} = 40.7^{+7.96}_{-7.49} & l = 69.9^{+12.1}_{-7.10} & b = 9.46^{+20.3}_{-16.5} \\
e_{c} = 4.00^{+1.64}_{-1.29} & l = 338^{+8.29}_{-6.27} & b = 12.8^{+5.86}_{-5.86} \nonumber 
\end{eqnarray}

\begin{center}
Galaxies $-$ Phoenix \& Leo I\ (n=8)
\end{center}
\begin{eqnarray}
e_{a} = 65.5^{+12.7}_{-13.6} & l = 281^{+33.3}_{-31.0} & b = -70.4^{+21.5}_{-4.57} \nonumber \\
e_{b} = 32.6^{+7.68}_{-10.9} & l = 69.0^{+14.1}_{-3.48} & b = -16.8^{+19.4}_{-24.0} \\
e_{c} = 3.94^{+2.18}_{-1.28} & l = 342^{+12.0}_{-3.34} & b = 9.85^{+5.41}_{-9.11} \nonumber 
\end{eqnarray}

The above solutions indicate that the outer satellites define a triaxial spatial distribution whose highly inclined long axis is nearly independent of the outer most object in the sample. These directions differ by
only $18^{\circ}\pm14$ [(1)$-$(2)] and $8.2^{\circ}\pm12$ [(1)$-$(3)]. The directions of the other principal axes are also quite similar. In spite of the fact that the number of objects outlining this structure is small, our working hypothesis is that the above result defines a real structure and in what follows we look for supporting evidence. We start by considering the spatial distribution of the 15 outer-most globular clusters and find:

\begin{center}
Globular Clusters\ (n=15)
\end{center}
\begin{eqnarray}
e_{a} = 24.7^{+8.38}_{-6.61} & l = 38.0^{+21.3}_{-18.6} & b = 47.4^{+14.5}_{-16.5} \nonumber \\
e_{b} = 20.1^{+6.00}_{-6.91} & l = 43.2^{+45.6}_{-26.1} & b = -42.4^{+19.9}_{-15.2} \\
e_{c} = 11.3^{+2.07}_{-2.60} & l = 311^{+16.1}_{-19.4} & b = -2.57^{+20.2}_{-18.1} \nonumber 
\end{eqnarray}

The above result suggests that there may be a real difference between the spatial distributions of the clusters and the galaxies. The cluster distribution appears to be more nearly oblate and not as flattened ($e_{c}/e_{a}$ = 0.46$^{+0.19}_{-0.18}$) as the galaxy distribution. The inner cutoff of $\sim25$ kpc in cluster selection is not a significant factor as very similar results are obtained if we omit the radial cutoff and instead choose clusters by their [Fe/H] values. For 99 clusters with [Fe/H] $< -1.0$ the shortest axis of the still nearly oblate spatial distribution is directed towards $l = 307^{+13.7}_{-13.8},~b = -3.31^{+12.7}_{-15.1}$ with axis ratio $e_{c}/e_{a}$ = 0.55$^{+0.15}_{-0.11}$.

 While both of the above samples are discouraging small, there is the strong suggestion that both luminous components of the outer halo of the Galaxy form structures whose minor axes are highly inclined to the Galactic rotation axis. This result is similar to that found in the seminal work of the mid 1970's. For comparison with this earlier work, the pole of the Magellanic Stream lies at $l = 11^{\circ}, b = -10^{\circ}$, (Lynden-Bell 1976) and the pole of the 12 members of the `Magellanic Plane Group' of Kunkel \& Demers (1976) is at $l = 347^{\circ}, b = 22^{\circ}$.

As a further check on our procedure we have examined the spatial distribution of the 34 metal-rich globular clusters with [Fe/H] $> -0.7$. The somewhat surprising results are: $e_{a} = 2.31^{+0.512}_{-0.538}$ towards $l = 347^{+13.6}_{-11.9},~b = -11.0^{+6.39}_{-3.55}$; $e_{b} = 1.13^{+0.296}_{-0.225}$ towards $l = 77.7^{+15.3}_{-12.1},~b = -1.89^{+8.27}_{-7.19}$ and $e_{c} = 0.343^{+0.109}_{-0.0734}$ towards $l = 357^{+28.8}_{-61.1},~b = 78.8^{+3.93}_{-7.76}$. A full discussion of these results, including observational selection effects, is beyond the scope of this paper, but we note that the short axis of the triaxial distribution now points toward high Galactic latitude while the long axis is only $\sim17^{\circ}$ off the Sun-center line which is very reminiscent of the bar-like feature believed to lie towards the Galactic center (c.f. \S 10.2 of Binney \& Merrifield 1998).

Before proceeding it is reasonable to consider how configurations such as we have seen in solutions (1),(2), and (3) could arise in practice. The possibility that the peculiar configuration observed has arisen by chance from what is actually a spherical distribution in the presence of the zone of avoidance was shown by a simple Monte Carlo simulation to have low probability (H96a). Two other possibilities are that we are seeing either the effects of dynamical evolution or a reflection of initial conditions. It is well known that dynamical friction will tend to bring satellites which are initially inclined to the plane of a galaxy into the plane working most efficiently on those objects closest to the plane initially (c.f. Quinn \& Goodman 1986). While this mechanism undoubtedly does operate in the inner part of the Galaxy where orbital times are relatively short, it is more  difficult to understand how it could significantly alter the distribution of the outer-most objects (Quinn \& Goodman 1986, Zaritsky et al. 1997). For now, we will consider the possibility that we may be seeing the effects of initial conditions, and after the following analysis of the radial velocities of the above samples will look for similar configurations in other nearby systems. 
  
\subsection{The Angular Variation of the Radial-Velocity Dispersion}
            
In order to look for possible kinematic signatures of the peculiar spatial distribution, radial-velocity dispersion tensors were computed in H96b for different samples of halo objects by the method of maximum likelihood. Here we confine our attention to the distant samples where the observed radial velocities are primarily a measure of the radial component of the velocity dispersion, and we use a mathematically simpler and equivalent procedure analogous to that used in determining the spatial distributions. First, the individual heliocentric radial velocities were corrected for the solar motion with respect to the local standard of rest, LSR, (using the value from Binney \& Merrifield 1998) and the rotation of the LSR with respect to the center of the Galaxy (assuming $V_{rot} = 220$ km~sec$^{-1}$). These radial velocities were then referred to the Galactic center by multiplying each by the secant of the angle at the object between the Galactic center and the Sun. The motion of the Galaxy with respect to the halo objects, $V_{G}$, was then determined by least squares. We define a radial-velocity dispersion tensor by computing the coefficients $\overline{V_{x}V_{x}}, \overline{V_{y}V_{y}}, \overline{V_{z}V_{z}}, \overline{V_{x}V_{y}}, \overline{V_{x}V_{z}}, \overline{V_{y}V_{z}}$ where $V_{x}= V_{r,G}\times \hat{x}$ etc  and $V_{r,G}$ is the observed radial velocity corrected for the two components of solar motion, referred to the Galaxy center and corrected for the above values of $V_{G}$. The direction cosines $\hat{x},\ \hat{y}$, and$\ \hat{z}$ are once again calculated with respect to the Galactic center. As the distance to the Galactic center is small compared with the Galactocentric distances of our objects, we do not expect much sensitivity to any tangential component of the velocity dispersion. We confirmed this by performing heliocentric solutions which give a different weighting to any tangential component and as expected found results almost identical to those below. Both methods yield large but similar values for $V_{G}$. As the direction of $V_{G}$ is usually close to being $90^{\circ}$ away from the major axis we are probably seeing the effects of rotation but given the extremely poor sensitivity to tangential motions, and the very small sample numbers such a motion will not be well determined. The radial-velocity dispersion tensor calculated as above was then diagonalized and the results are given below along with the corresponding value of $V_{G}$ in km~sec$^{-1}$. The units of the eigenvalues denoted by $\sigma$ are km~sec$^{-1}$.

\begin{center}
All Galaxies\ (n=10)
\end{center}
\begin{eqnarray}
\sigma_{a} = 52.6^{+6.23}_{-22.8} & l = 326^{+67.7}_{-54.9} & b = -78.4^{+22.0}_{-0.840} \nonumber \\
\sigma_{b} = 35.2^{+0.390}_{-24.8} & l = 69.5^{+14.0}_{-7.92} & b = -2.68^{+24.6}_{-20.7} \\
\sigma_{c} = 8.12^{+1.21}_{-6.10} & l = 340^{+10.2}_{-7.08} & b = 11.3^{+3.61}_{-7.55} \nonumber \\ 
V_{G} = 152_{-27.8}^{+137} & l = 112_{-57.9}^{+42.1} & b = -17.9_{-4.05}^{+18.5} \nonumber
\end{eqnarray}

\begin{center}
Galaxies $-$ Phoenix\ (n=9)
\end{center}
\begin{eqnarray}
\sigma_{a} = 47.0^{+3.05}_{-26.5} & l = 3.80^{+46.9}_{-83.1} & b = -76.8^{+28.1}_{-3.50} \nonumber \\
\sigma_{b} = 37.8^{+1.13}_{-26.4} & l = 70.7^{+16.1}_{-10.5} & b = 5.25^{+28.6}_{-17.0} \\
\sigma_{c} = 8.71^{+1.02}_{-7.18} & l = 339^{+7.65}_{-11.4} & b = 12.1^{+8.57}_{-7.52} \nonumber \\
V_{G} = 135_{-15.4}^{+114} & l = 100_{-60.8}^{+45.8} & b = -11.3_{-7.89}^{+20.2} \nonumber
\end{eqnarray}

\begin{center}
Galaxies $-$ Phoenix \& Leo I\ (n=8)
\end{center}
\begin{eqnarray}
\sigma_{a} = 33.7^{+0.639}_{-19.4} & l = 268^{+14.1}_{-5.88} & b = -62.9^{+35.6}_{-16.9} \nonumber \\
\sigma_{b} = 23.2^{+0.676}_{-19.1} & l = 74.3^{+3.25}_{-23.0} & b = -26.5^{+17.9}_{-33.5} \\
\sigma_{c} = 3.70^{+0.276}_{-2.84} & l = 347^{+5.98}_{-6.12} & b = 5.45^{+9.94}_{-2.10} \nonumber \\ 
V_{G} = 201_{-48.1}^{+84.9} & l = 142_{-20.6}^{+13.5} & b = -5.35_{-8.16}^{+11.3} \nonumber
\end{eqnarray}

\begin{center}
Globular Clusters\ (n=15)
\end{center}
\begin{eqnarray}
\sigma_{a} = 81.8^{+3.18}_{-34.4} & l = 3.89^{+15.9}_{-26.8} & b = 25.7^{+20.4}_{-4.62} \nonumber \\
\sigma_{b} = 79.7^{+3.22}_{-54.2} & l = 80.9^{+4.30}_{-33.6} & b = -25.1^{+20.8}_{-15.5} \\
\sigma_{c} = 31.4^{+6.23}_{-12.3} & l = 313^{+14.8}_{-25.8} & b = -52.7^{+26.7}_{-4.40} \nonumber \\ 
V_{G} = 96.2_{-170}^{+105} & l = 270_{-21.8}^{+59.9} & b = -17.4_{-5.00}^{+52.3} \nonumber
\end{eqnarray}

As can be seen from above, all three galaxy solutions are nearly perfectly aligned with the corresponding spatial solutions of the previous section and in the sense that the largest dispersion axis coincides with the longest spatial axis and vice versa. We cautiously interpret this result as support for our hypothesis. That this is not the case for the outer clusters, however, is another indication that the cluster component is different from the galaxy component. It should emphasized that what is being measured above is the radial-velocity dispersion and that without knowledge of the tangential component of the velocity dispersion the kinematical state of the outer halo remains incomplete.

\section{The Spatial Distribution of Satellites in the M~31 Sub-Group}

If our hypothesis is correct then we might expect to see similar structures around other galaxies. M~31 would appear to be an ideal test case as it has at least 11 satellite galaxies with reasonably well determined distances according to the recent review of Mateo (1998). The galaxy EGB0427+63 was not included in our sample due to the large quoted uncertainty in its distance and an updated distance of 660 kpc was assumed for IC10 (Sakai et al. 1999).  The spatial distribution of the 11 members of this grouping of galaxies as denoted by Mateo(1998) (also see Karachentsev, 1996) was determined below using identical procedures as in \S 2.2 except that $d_{i}$ is now taken to be the true distance from M~31. As before, the units of $e_{a,b,c}$ are kpc.
  
\begin{center}
M~31 Sub-group\ (n=11)
\end{center}
\begin{eqnarray}
e_{a} = 110^{+42.1}_{-30.1} & l = 270^{+20.2}_{-32.1} & b = -63.9^{+20.2}_{-11.6} \nonumber \\
e_{b} = 63.4^{+20.4}_{-25.7} & l = 299^{+9.39}_{-8.82} & b = 23.2^{+20.3}_{-20.7} \\
e_{c} = 8.83^{+5.89}_{-5.04} & l = 24.1^{+6.10}_{-4.32} & b = -11.4^{+5.99}_{-6.45} \nonumber 
\end{eqnarray}

Noting that the distance errors for the M~31 companions are generally larger than those for the Galaxy and that the low Galactic latitude of M~31 almost certainly contributes to the incompleteness of the sample, the results in (9) are quite remarkable not only because they suggest a similar triaxial structure around M~31 but that the long axis of this structure is within $16^{\circ}\pm17$ of that found for the Galactic satellites in solution (1). A solution combining both the 10 Galaxy companions and the M~31 satellites translated to a common origin is given below.
 
\begin{center}
Galaxy \& M~31 Satellites Combined\ (n=21)
\end{center}
\begin{eqnarray}
e_{a} = 107^{+21.8}_{-16.2} & l = 276^{+12.2}_{-30.2} & b = -70.6^{+11.4}_{-8.47} \nonumber \\
e_{b} = 45.2^{+10.8}_{-14.2} & l = 278^{+13.2}_{-12.2} & b = 19.4^{+10.6}_{-11.4} \\
e_{c} = 15.4^{+4.71}_{-4.62} & l = 7.75^{+10.4}_{-12.6} & b = -0.640^{+5.67}_{-7.56} \nonumber 
\end{eqnarray}

Figure 1 shows a projection of the 3-dimensional representation of this combined sample. As an aside one could also consider Figure 1 to be a visual illustration of the fact that the ratio of the density of luminous matter to the total matter density i.e. $\Omega_{\star}/\Omega_{0}\sim 0.01$ when one realizes that the pile of lighted baryons which we call the Galaxy is contained well within the elliptical contour at the origin.

\vspace{0.25in}
Fig. 1 here
\vspace{0.25in}
 
\section{A Relationship with Larger Scale Structure}

Given the results of the previous sections we are encouraged to trace the signature of solutions (1), (2), and (3) to larger scales. We start by considering the spatial distribution of the isolated members of the Local Group, then consider the kinematics of the high-velocity clouds which are potential Local Group interlopers and finally look at the distribution of galaxies in the Coma-Sculptor cloud, a nearby large-scale structure in which the Local Group is embedded.
 
\subsection{The Spatial distribution of Local Group Galaxies}

Excellent reviews of the properties of Local Group galaxies have recently been given by Mateo (1998) and van den Bergh (2000). For our purposes we focus attention on the spatial distribution of a subset of these galaxies and consider only those galaxies which are relatively isolated (principally the dwarf irregulars, i.e. those galaxies which are not obvious satellites of/nor either the Galaxy or M~31). Given that this gas-rich subset likely contains potential but as yet `unincorporated' galaxy building blocks, we argue that there is the reasonable possibility that the spatial distribution of these galaxies could provide us with an outline of the original shape of the Local Group. Mateo (1998) argues the case for a more extended Local Group membership than van den Bergh, and based on the discussion in Mateo (1998) we have chosen our sample from his compilation. It includes: WLM, NGC55, IC1613, Leo A, NGC3109, GR8, SagDIG, NGC6822, DDO210, IC5152, Tucana, UKS2323-326, and Pegasus. Mateo's (1998) distances were used for each of the objects. The center of mass was assumed to be at 460 kpc from the Galaxy in the direction of M~31. The procedure followed is identical to that used to determine the spatial distribution of Galactic satellites, and the results are given below. The units of $e_{a,b,c}$ are kpc.

\begin{center}
Local Group (Isolated Dwarfs)\ (n=13)
\end{center}
\begin{eqnarray}
e_{a} = 654^{+145}_{-88.7} & l = 349^{+22.2}_{-15.2} & b = -54.7^{+11.3}_{-8.83} \nonumber \\
e_{b} = 335^{+69.5}_{-120} & l = 307^{+21.3}_{-21.1} & b = 28.0^{+12.8}_{-14.4} \\
e_{c} = 128^{+25.5}_{-42.6} & l = 48.2^{+20.1}_{-16.3} & b = 19.7^{+7.44}_{-13.4} \nonumber 
\end{eqnarray}

\noindent
In spite of the larger uncertainties in the distances of many Local Group galaxies, we note the similarity of the above distribution to solutions (1), (2), and (3) for the Galaxy (except for a reversal of the intermediate and smallest axes). Had we chosen our subset from van den Bergh's (2000) compilation, NGC55, NGC3109, GR8, IC5152, and UKS2323-326 would be deleted from the list above. A solution with the remaining 8 galaxies also shows a triaxial distribution with the long axis at $l = 358^{+20.2}_{-11.4},\ b = -34.1^{+13.4}_{-20.9}$. Due to uncertainties both in defining the zero-velocity surface, which leads to the definition of Local Group membership, and in defining the `isolated, gas-rich' subset which by necessity involves small numbers, we are prompted to look beyond the Local Group and consider the spatial distribution of the galaxies making up the nearest large-scale structure. Before doing so, we examine the kinematics of the high-velocity clouds.
 
\subsection{Kinematics of High-Velocity Clouds}

It is clear that additional probes of the region occupied by the satellite galaxies are desirable in order to strengthen the hypothesis. Recent work on high-velocity clouds (HVC's) attempts to make the case for them being Local Group interlopers (c.f. Blitz et al. 1999, Braun \& Burton 1999). Solar motion solutions by both groups of authors reveal a residual which is interpreted as a systemic infall towards the Sun (c.f. Figure 5 of Braun \& Burton (1999)). Below we examine the angular variation in the velocity field by evaluating the quadrupolar term using the isolated, compact (but incomplete) sample of Braun \& Burton. The starting point is the following expression for the radial-velocity component of the velocity field expanded about the origin (the Sun in this case) for an individual HVC

\begin{eqnarray}
v_{r}=\bf{n \cdot v_{0} + n\cdot D \cdot r}
\end{eqnarray}
\noindent
where $\bf{n}=\bf{r}/|r|$ is the unit vector containing the object's direction cosines, $-v_{0}$ is the solar motion, $\bf{D}$ is the displacement tensor (c.f. eqn 2.72 of Ogorodnikov, 1965) and $\bf{r}$ is the position vector. Since only the directions and not the distances are known for the clouds, we replace $\bf{r}$ with $\bf{n}$ in the above equation and solve for the components of $\bf{v_{0}}$ and the symmetrical anisotropy tensor by least squares. Eigenvalues are in km~sec$^{-1}$.

\begin{center}
Velocity Anisotropy Tensor for HVC's\ (n=66)
\end{center}
\begin{eqnarray}
e_{a} = 5.56^{+12.5}_{-24.4} & l = 352^{+32.2}_{-29.5} & b = -61.6^{+19.5}_{-5.77} \nonumber \\
e_{b} = 64.1^{+29.6}_{-10.5} & l = 66.6^{+9.96}_{-11.7} & b = 8.02^{+11.3}_{-16.9} \\
e_{c} = -69.6^{+13.6}_{-23.0} & l = 332^{+7.11}_{-8.83} & b = 27.0^{+13.4}_{-12.1} \nonumber \\ 
V_{\odot} = 278_{-13.6}^{+13.3} & l = 83.9_{-3.50}^{+3.24} & b = -15.2_{-4.17}^{+5.79} \nonumber
\end{eqnarray}

The monopole term is $V_{mono}=-105^{+8.39}_{-9.30}$ km~sec$^{-1}$ and this value (the average of the original 3 eigenvalues) has been subtracted from the original 3 eigenvalues to give the values tabulated above. Once again one can  identify the directions in the above solution with those of (1), (2), and (3). This identification should be considered tentative until the sample is complete but it should also provide even more incentive to determine the true nature (at least the distances) of the HVC's.  
 
\subsection{Nearby Large-Scale Structure}

We note that the shortest axis of the Local Group spatial distribution lies within $13^{\circ}\pm10$ of the direction of the supergalactic pole at $l = 47.4^{\circ}$ and $b = + 6.32^{\circ}$, suggesting a possible connection with nearby large-scale structure. Examination of the Nearby Galaxies Atlas (Tully \& Fisher 1987) shows that the Local Group is embedded in a filament which the above authors refer to as the Coma-Sculptor Cloud. The spatial distribution of the galaxies in the Coma-Sculptor Cloud was quantified in the same manner described above. A solution was carried out for those galaxies with a group prefix designation of 14 from Tully's (1988) catalog and is given below. The center of gravity of the 214 galaxies was determined by giving each galaxy unit weight and was found to lie at 2.8 h$^{-1}$Mpc towards l = 155$^{\circ}$, b = 83.5$^{\circ}$. The units of $e_{a,b,c}$ below are h$^{-1}$Mpc with h = H$_{0}$/100.
        
\begin{center}
Coma-Sculptor Cloud\ (n=214)
\end{center}
\begin{eqnarray}
e_{a} = 2.25^{+0.0955}_{-0.0980} & l = 316^{+5.48}_{-3.75} & b = -78.0^{+3.26}_{-2.37} \nonumber \\
e_{b} = 0.769^{+0.076}_{-0.0874} & l = 307^{+1.84}_{-2.71} & b = 11.8^{+3.37}_{-2.50} \\
e_{c} = 0.127^{+0.0159}_{-0.0085} & l = 37.4^{+2.06}_{-2.78} & b = 1.92^{+0.991}_{-0.882} \nonumber 
\end{eqnarray}

\noindent
Once again we note that the long axis of the distribution is pointing at high
Galactic latitude while the short axis differs from the position of the supergalactic pole by only $11^{\circ}$.

\subsection{Tidal Effects}

Since gravity is responsible for the growth of structure in the universe it is relevant to consider the effects of the quadrupolar (tidal) component of the gravitational field on the structures we have discussed above. Following Raychaudhury \& Lynden-Bell (1989) (RLB89), we define the tidal tensor as
\begin{eqnarray}
Q_{i,j}=\sum_{\alpha}\frac{GM_{\alpha}}{d_{\alpha}^{3}}[t_{i,j}]
\end{eqnarray}
where the tensor components $t_{1,1}=\frac{3~x_{\alpha}x_{\alpha}}{d^{2}_{\alpha}}- 1, t_{1,2}=\frac{3~x_{\alpha}y_{\alpha}}{d^{2}_{\alpha}}, t_{1,3}=\frac{3~x_{\alpha}z_{\alpha}}{d^{2}_{\alpha}}$ etc and as before $d_{\alpha}$ is the distance of the $\alpha^{th}$ object from the origin under consideration and $x_{\alpha},y_{\alpha}$ and $z_{\alpha}$ are the components of $d_{\alpha}$ projected onto the principal axes of the Galactic coordinate system. $M_{\alpha}$ is the mass of the ${\alpha}^{th}$ galaxy. In computing $Q_{i,j}$ we have used reddening corrected apparent B magnitudes (where available) and assumed $M/L_{B}=20$ to obtain $M_{\alpha}/d^{2}_{\alpha}$ and distances from the Tully (1988) catalog scaled to $H_{0}=100$ km~sec$^{-1}$Mpc$^{-1}$ for galaxies beyond the Local Group and the Mateo (1998) compilation for distances within the Local Group. For comparison with the results of RLB89 the units of the eigenvalues of Q given below are $G\times(2.32\times10^{12} M_{\odot})/(Mpc^{3})$. First, using the Local Group barycenter as origin we calculate Q for the effects of galaxies in the Coma-Sculptor cloud (as defined in \S 3.2) on the Local Group. The results are:

\begin{center}
Tidal Quadrupole$-$(Coma-Sculptor Cloud on Local Group)\ (n=176)
\end{center}
\begin{eqnarray}
e_{a} = 0.0944^{+0.0243}_{-0.0376} & l = 319^{+4.85}_{-8.19} & b = -51.3^{+13.5}_{-12.7} \nonumber \\
e_{b} = -0.0122^{+0.0165}_{-0.0139} & l = 310^{+3.54}_{-5.37} & b = 38.2^{+13.2}_{-13.5} \\
e_{c} = -0.0822^{+0.0284}_{-0.0157} & l = 43.3^{+2.96}_{-5.85} & b = 4.60^{+1.99}_{-4.16} \nonumber 
\end{eqnarray}

Comparing these results with RLB89 we find excellent agreement both in relative amplitude, sign, and orientation of the eigenvalues. We note that the positive eigenvalue (tidal stretch) is in a direction in good agreement with our calculated long axis of the spatial spatial distribution of Local Group galaxies in \S 4.1 while the directions of the negative eigenvalues (tidal compression) are easily identified with the previously determined short and intermediate axes. Thus these tides are acting to maintain the observed Local Group spatial distribution.

If we now calculate the tidal effects at the Galactic center due to the Coma-Sculptor cloud and Local Group galaxies except the Galaxy and its companions we find
 
\begin{center}
Tidal Quadrupole$-$(Surrounding galaxies on the Milky Way)\ (n=190)
\end{center}
\begin{eqnarray}
e_{a} = 1.48^{+1.30}_{-1.32} & l = 303^{+7.60}_{-0.949} & b = 21.4^{+2.21}_{-1.71} \nonumber \\
e_{b} = -0.654^{+0.662}_{-0.665} & l = 327^{+13.4}_{-14.0} & b = -66.7^{+5.04}_{-1.56} \\
e_{c} = -0.825^{+0.663}_{-0.639} & l = 36.6^{+5.78}_{-1.64} & b = 8.76^{+5.05}_{-5.50} \nonumber 
\end{eqnarray}

While the directions are similar to those in (1), (2), and (3), the amplitudes and signs are such that the largest positive eigenvalue (tidal stretch) is in the direction of M~31 which generally corresponds with the shortest spatial axis in \S 2.2. Similarly we find tidal compression indicated along our longest spatial axes. These effects would disrupt the observed satellite distribution if the tidal forces dominate. However, with M$_{B,Galaxy}$= $-$20.5 (Tully 1988) and for consistency, the above assumed $M/L_{B}=20$, our Galaxy would have a mass of $0.46\times10^{12} M_{\odot}$.  Assuming a radius of $0.2$ Mpc, the value of $GM/R^{3}$ in the above units for our Galaxy is then $\sim17$ times the largest eigenvalue in (17), and we conclude that at least for now the Galaxy is dense enough to withstand the tidal forces exerted on it by its neighbors.
 
\section{Discussion}

Our working hypothesis is that the spatial distribution of the companion galaxies of the Milky Way defines a real triaxial structure. As a measure of the degree of support for this hypothesis we identify and compute the difference in the direction of the long axis of the possibly related distributions from a given reference value. For this reference we use the long axis of the Coma-Sculptor spatial distribution in solution (14) and give the difference for each solution- a) for the spatial distribution of Galaxy satellites (solutions (1), (2), and (3)) these differences are $3.6^{\circ}\pm9.4$, $14^{\circ}\pm12$, and $12^{\circ}\pm12$ respectively. b) the angular variation of radial-velocity dispersion of Galaxy satellites ((5), (6), and (7)) give differences of $2.1^{\circ}\pm12$, $10^{\circ}\pm13$, and $21^{\circ}\pm2.3$. c) for the M~31 sub-group and M~31 and Galaxy satellites combined ((9) and (10)) the differences are $20^{\circ}\pm5$, and $13^{\circ}\pm9.1$. d) for the isolated dwarfs of the Local Group (solution (11)) the difference is $26^{\circ}\pm10$. Finally for the velocity anisotropy tensor of the high-velocity clouds, solution (13), we find a difference of $20^{\circ}\pm5.6$. In addition, we note that the shortest axes of both the Local Group distribution and the Coma-Sculptor cloud differ by 13$^{\circ}\pm10$ and 11$^{\circ}\pm2.3$ from the supergalactic pole. Tidal quadrupoles, (16) and (17) also exhibit close alignment with both of the above principal directions with the tidal effect of the Coma-Sculptor galaxies acting to maintain the Local Group structure, while the tidal effects of M~31 and surroundings are not sufficiently strong to disrupt the present structure surrounding the Galaxy. The outer globular clusters do not appear to be part of this same structure although they do form a nearly oblate, flattened system with minor axis highly inclined to the present rotation axis. 

It is the concordance of the directions of these principal axes which leads us to conclude that the above interpretation of what we observe in the outer regions of the Galaxy is correct, and that it is a reflection of structure on much larger scales. Given the above scenario we are left with having to explain why none of the principal axes is aligned with the Galactic pole. Modelling of galaxy formation in CDM cosmologies shows that the orientation of the axis of the resulting gaseous disk may bear little relation to the original rotation axis of the proto-galaxy because angular momentum is transfered from the inner gaseous disk to the dark matter dominated outer regions during the gravitational interaction with the dark matter clumps as they merge together. Katz \& Gunn (1991) find, for example, a deviation of $\sim 30^{\circ}$ between disk axis and rotation axis of the original distribution of lumps. In our case this misalignment angle is $\sim12-30^{\circ}$ assuming that the original rotation axis is the major axis. Thus we have a picture that appears plausible though somewhat difficult to falsify due to the small number of satellites. However, the possibility that the distribution of dark matter in the Galaxy may be more like that of the outer satellites is potentially testable. Further, there may already be some observational precedents. In addition to anticipating the results found here for the Milky Way, Holmberg (1969) showed on the basis of visual inspection of Palomar Sky Survey plates that galaxy satellites within $\sim$ 40~h$^{-1}$ kpc of many parent spirals have a tendency to be located along the disk minor axis. Further, Zaritsky et al. (1997) have shown that the observed distribution of satellite galaxies of spiral primaries out to $\sim$ 375~h$^{-1}$ kpc is asymmetric and is also elongated along the disk minor axis.

Given the apparent ubiquitousness of this `Holmberg' effect combined with what we have found above, the following galaxy formation scenario suggests itself. Early on, a preferred axis is established in the direction of dominant motions (such as along a filament as above) and this axis also becomes the rotation axis (this is certainly not unambiguously determined here, but our results for $V_{G}$ are consistent with this idea). This axis also serves as a conduit along which the bulk of the protogalactic lumps are and may still be accreted. The Katz and Gunn (KG) mechanism then causes a misalignment between this axis and the axis of the gaseous disk as it forms by dissipation at the center. The KG mechanism would ensure the observed `stochasticity' in the degree of misalignment. Evidence for this galaxy formation scheme should be revealed in the increasingly sophisticated numerical simulations of structure formation.

\acknowledgments

The author wishes to thank Dr. Mike Hudson for many useful discussions and suggestions on this subject, the anonymous referee for constructive comments, and Stephen Gwyn for help with the figure.  He is also wishes to acknowledge financial support from an NSERC of Canada operating grant.

\clearpage

\begin{figure}
\plotone{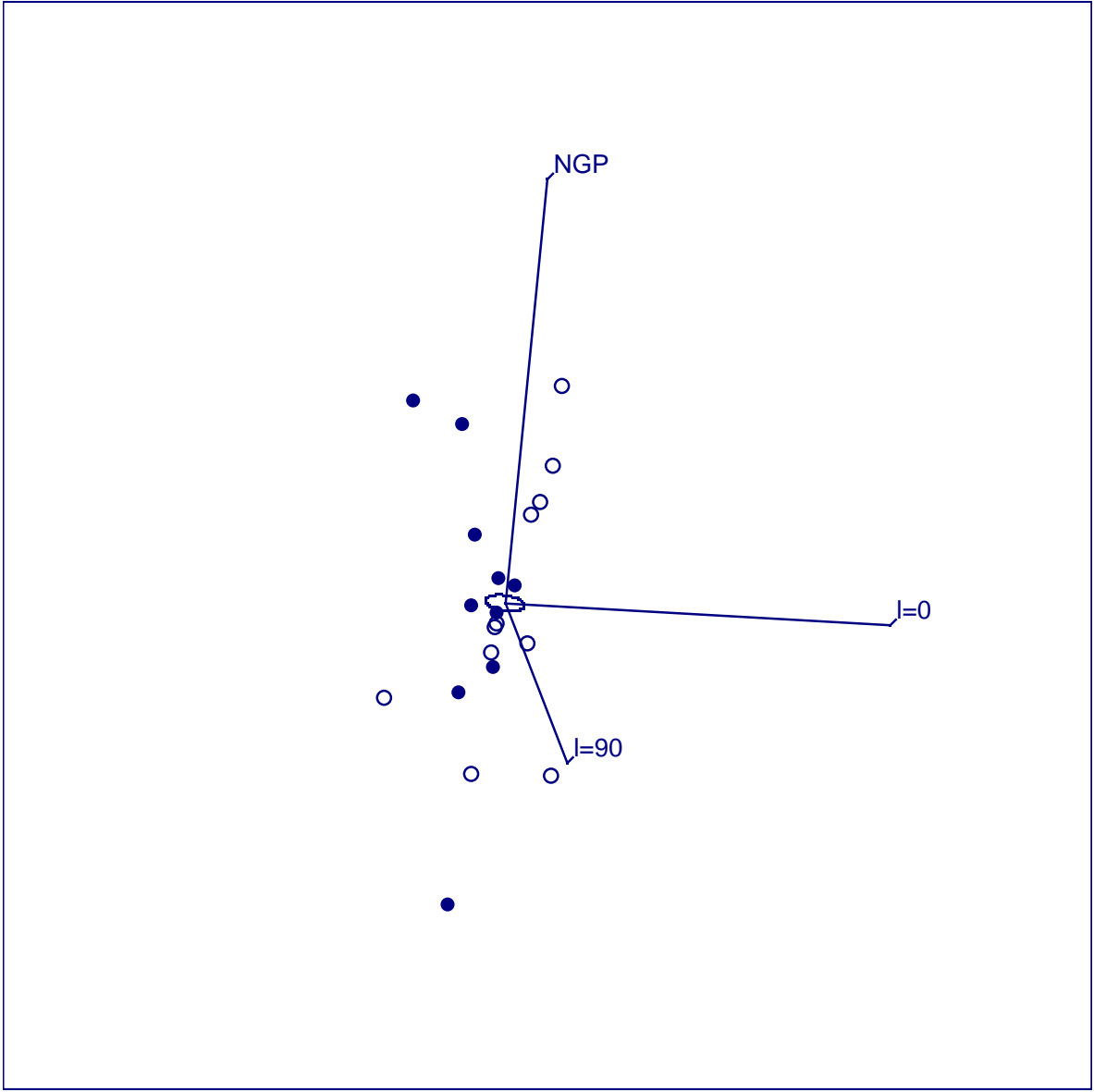}
\figcaption{
The spatial distribution of 10 Galactic satellites (closed circles) and 11 M~31 satellites translated to the Galactic origin (open circles) plotted in Galactic coordinates. The axes are 500 kpc in length, and the ellipse at the center shows the distortion due to projection of a circle of radius 25 kpc in the Galactic plane. The quantitative description of this distribution is given in (10).}

\end{figure}
  
\end{document}